\documentclass[superscriptaddress,aps,showpacs]{revtex4}

\usepackage{amsmath,amsbsy,amssymb}

\newcommand{\bd}[2]{#1\!\cdot\!#2}
\newcommand{\bphi}{\boldsymbol\phi}
\newcommand{\bpsi}{\boldsymbol\psi}
\newcommand{\bchi}{\boldsymbol\chi}
\newcommand{\bphib}{\bar{\bphi}}
\newcommand{\bpsib}{\bar{\bpsi}}
\newcommand{\bchib}{\bar{\bchi}}
\newcommand{\Sch}{Schr\"odinger }
\newcommand{\Pai}{Painlev\'e }

\begin{document}

\title{The Sasa-Satsuma higher order nonlinear \Sch equation\\ and its
  bilinearization and multi-soliton solutions} 
\author{C. Gilson}\email{c.gilson@maths.gla.ac.uk}
\affiliation{Department of Mathematics, University of Glasgow, Glasgow
  G12 8QW, UK}
\author{J.  Hietarinta}\email{jarmo.hietarinta@utu.fi}
\affiliation{Department of Physics, University of Turku, FIN-20014
  Turku, Finland}
\author{J.  Nimmo}\email{j.nimmo@maths.gla.ac.uk}
\affiliation{Department of Mathematics, University of Glasgow, Glasgow
  G12 8QW, UK}
\author{Y. Ohta}\email{ohta@math.kobe-u.ac.jp}
\affiliation{Department of Mathematics, Kobe University, Rokko, Kobe
  657-8501, Japan}

\begin{abstract}{Higher order and multicomponent generalizations of the
    nonlinear \Sch equation are important in various applications,
    e.g., in optics. One of these equations, the integrable
    Sasa-Satsuma equation, has particularly interesting soliton
    solutions. Unfortunately the construction of multi-soliton
    solutions to this equation presents difficulties due to its
    complicated bilinearization. We discuss briefly some previous
    attempts and then give the correct bilinearization based on the
    interpretation of the Sasa-Satsuma equation as a reduction of the
    three-component Kadomtsev-Petviashvili hierarchy.  In the process
    we also get bilinearizations and multi-soliton formulae for a two
    component generalization of the Sasa-Satsuma equation (the
    Yajima-Oikawa-Tasgal-Potasek model), and for a $(2+1)$-dimensional
    generalization.}
\end{abstract}

\maketitle

\section{Introduction}
One of the most interesting applications of solitons is in the
propagation of short pulses in optical fibers (for an overview, see
\cite{KA}). The basic phenomena are described by the nonlinear \Sch
equation, but as the pulses get shorter various additional effects
become important. In \cite{KH} Kodama and Hasegawa derived the
relevant equation with higher order correction terms, the generic form
of such an equation is (in the optical fiber setting the roles of time
and space are usually reversed)
\begin{equation}
\label{E:generic}
  iq_\xi+\alpha_1q_{\tau\tau}+\alpha_2 |q|^2q + i[\beta_1
q_{\tau\tau\tau}+\beta_2 |q|^2 q_\tau
   + \beta_3 q(|q|^2)_\tau]=0,
\end{equation}
where the $\alpha_i,\beta_i$ are real constants and $q$ a complex function.
The first three terms form the standard nonlinear \Sch equation (nls) and the
$\beta_i$ terms are the perturbative corrections.  Usually one chooses the
scaling so that $\alpha_2=2\alpha_1$. In this paper we assume $\beta_1\neq 0$.

Our main concern here is the bilinearization and multi-soliton
solutions of the Sasa-Satsuma equation (SSnls) \cite{SS}, which is a
particularly interesting integrable example in the above class. In
this introductory section we discuss some basic properties of
\eqref{E:generic}, its integrable special cases and their
multicomponent generalizations. In particular we show that many
previous attempts to solve these equations have produced only rather
trivial solutions, in which the complex and multicomponent freedom has
been ``frozen''.  The reason for this turns out to be in the incorrect
bilinearization that was used in those papers.  The correct
bilinearization (presented in Sec.~\ref{Sec2} with detailed derivation
in Sec.~\ref{Sec3}) follows once we identify SSnls as a reduction of
the three-component Kadomtsev-Petviashvili (KP) hierarchy, and then we
also obtain general multi-soliton solutions.

\subsection{Gauge transformation} 
In order to understand the complex structure of (\ref{E:generic}) it
is important to isolate the gauge (phase) invariance and fix the
gauge.  First let us recall that the nls part of (\ref{E:generic})
(i.e., if $\beta_i=0$) is invariant under the combined gauge/Galilei
transformation
\begin{equation}
q(\xi,\tau)=e^{iv(\tau -v\xi )/\alpha_1}y(x,t),\quad x=\tau -2v\xi
,\,t=\xi.\label{nlsgau}
\end{equation}
The full equation (\ref{E:generic}) is not invariant under
\eqref{nlsgau}, but if we try the transformation
\begin{equation}\label{E:gaugetra}
q(\xi,\tau)=e^{i(c_1\tau +c_2\xi )}y(x,t),\quad x=\tau +c_3 \xi
,\,t=\xi,\quad c_i \text{ real constants,}
\end{equation}
we find that if
\begin{equation}
c_3=c_1(-2\alpha_1+3\beta_1c_1),\quad
c_2=c_1^2(-\alpha_1+c_1\beta_1),
\end{equation}
then (\ref{E:generic}) is {\em form invariant}: the equation for $y(x,t)$ is
as in (\ref{E:generic}) with $\beta_i$ unchanged, but the $\alpha_i$ change
according to
\begin{equation}\label{E:gaugepar}
\alpha_1\to\widetilde\alpha_1=\alpha_1-3\beta_1c_1,\quad
\alpha_2\to\widetilde\alpha_2=\alpha_2-\beta_2c_1.
\end{equation}
We can therefore use this transformation to put $\alpha_1=
\alpha_2=0$,
provided that
\begin{equation}\label{E:gaugecond}
3\beta_1\alpha_2=\beta_2\alpha_1.
\end{equation}
[In the usual normalization $\alpha_1=\frac12, \alpha_2=1$
(\ref{E:gaugecond}) means $\beta_2=6\beta_1$.] In all integrable cases
(along with some non-integrable cases appearing in the literature)
(\ref{E:gaugecond}) is satisfied, and we assume it from now on.

On the basis of the above {\em we fix the gauge \eqref{E:gaugetra} by
  requiring that $\alpha_i=0$ in the equation}, and compare results
only in that uniquely defined gauge.

\subsection{Integrable cases} The integrability of the class of equations
(\ref{E:generic}) has been studied by a number of authors using \Pai
analysis \cite{GSB97,MTC,Sa97} and other methods \cite{NR}, with the consistent
result that if $\beta_1,\beta_2\neq 0$ there are precisely two integrable
cases with bright solitons:
\begin{enumerate}
\item Hirota's equation (Hnls) \cite{H73}:
  $\beta_1:\beta_2:\beta_3=1:6:0$,
\begin{equation}
q_t+q_{xxx}+ 6|q|^2q_x=0, \label{E:H}
\end{equation}
\item Sasa-Satsuma equation (SSnls) \cite{SS}: $\beta_1:\beta_2:\beta_3=1:6:3$,
\begin{equation}
q_t+q_{xxx}+6|q|^2 q_x + 3q |q^2|_x=0.\label{E:SS}
\end{equation}
\end{enumerate}
Here the scaling convention mentioned above has been assumed and the
$\alpha_i$ terms eliminated.

Some non-integrable special cases of \eqref{E:generic} have also been
studied in the literature, including \cite{LW,RL96}:
$\beta_1:\beta_2:\beta_3=1:6:6$,
\begin{equation}
q_t+q_{xxx}+ 6(q |q^2|)_x=0.\label{E:nonint}
\end{equation}

\subsection{Multicomponent generalizations}
Both Hnls and SSnls allow various kinds of multicomponent generalizations,
some of them integrable. The results of a \Pai analysis \cite{ST} can be
summarized as follows:
\begin{equation}
\text{Case }1:\quad\left\{\begin{array}{rcl}
q_{1t}+q_{1xxx}+3(|q_1|^2+|q_2|^2)q_{1x}&=&0,\\
q_{2t}+q_{2xxx}+3(|q_1|^2+|q_2|^2)q_{2x}&=&0,
\end{array}\right.\label{E:sako3}
\end{equation}
which can be interpreted as a real 4-component modified
Korteweg--de~Vries (mKdV) equuation, reducing to Hnls for $q_1=q_2$
etc.,
\begin{equation}
\text{Case 2 \cite{ST}}:\quad\left\{\begin{array}{rcl}
q_{1t}+q_{1xxx}+3(|q_1|^2+|q_2|^2)q_{1x}+3q_1(|q_2|^2)_x&=&0,\\
q_{2t}+q_{2xxx}+3(|q_1|^2+|q_2|^2)q_{2x}+3q_2(|q_1|^2)_x&=&0,
\end{array}\right.\label{E:sako1}
\end{equation} and
\begin{equation}
\text{Case 3 \cite{PSM94}}:\quad\left\{\begin{array}{rcl}
q_{1t}+q_{1xxx}+3(|q_1|^2+|q_2|^2)q_{1x}+\tfrac32q_1(|q_1|^2+|q_2|^2)_x&=&0,\\
q_{2t}+q_{2xxx}+3(|q_1|^2+|q_2|^2)q_{2x}+\tfrac32q_2(|q_1|^2+|q_2|^2)_x&=&0,
\end{array}\right.\label{E:sako5}
\end{equation}
which reduce to SSnls under the above reduction, and a mixed case
\begin{equation}
\text{Case 4 \cite{YO,TP}}:\quad\left\{\begin{array}{rcl}
q_{1t}+q_{1xxx}+\tfrac{a}2(|q_1|^2+|q_2|^2)q_{1x}+
\tfrac{a}2q_1(q_1^*q_{1x}+q_2^*q_{2x})&=&0,\\
q_{2t}+q_{2xxx}+\tfrac{a}2(|q_1|^2+|q_2|^2)q_{2x}+
\tfrac{a}2q_2(q_1^*q_{1x}+q_2^*q_{2x})&=&0,\\
\end{array}\right.\label{E:sako4}
\end{equation}
which reduces to Hnls under $q=q_1=q_2,a=3$ and to SSnls under
$q=q_1=q_2^*,a=6$. In each case we must of course adjoin the complex
conjugated equations. Cases 1--3 are invariant under
$q_2\leftrightarrow q_2^*$, while case 4 changes to the alternative form
\begin{equation}
\text{Case }4':\quad\left\{\begin{array}{rcl}
q_{1t}+q_{1xxx}+\tfrac{a}2(|q_1|^2+|q_2|^2)q_{1x}+
\tfrac{a}2q_1(q_1^*q_{1x}+q_2q_{2x}^*)&=&0,\\
q_{2t}+q_{2xxx}+\tfrac{a}2(|q_1|^2+|q_2|^2)q_{2x}+
\tfrac{a}2q_2(q_1q_{1x}^*+q_2^*q_{2x})&=&0.\\
\end{array}\right.\label{E:sako2}
\end{equation}
Under the reduction $q_2=0$ cases 1,2,4 reduce to Hnls, and 3 to SSnls.

\subsection{The mKdV limit}
With complex and multicomponent equations it is important to make the
following observation: we can always make the real, one-component reduction
\begin{equation}
q_i(x,t)=c_i\,u(x,t),\,\forall i,\label{E:realr}
\end{equation}
where $u$ is a {\em real} function and $c_i$ are arbitrary complex
constants. As a result of this {\em all} the equations mentioned
before (and many others, including non-integrable ones) reduce to the
real mKdV equation 
\begin{equation}
u_t=u_{xxx}+\kappa u^2 u_x.
\end{equation}
(Note that for case 2 we need $|c_1|=|c_2|$.)
This was observed already in \cite{GSB97}, see eqs (21-25).  A
consequence of this rather simple observation is the following:

\vskip 0.2cm
\noindent{\em 
  All these complex and/or multicomponent systems always have multi-soliton
  solutions of the real mKdV type, with frozen complex and/or multicomponent
  freedom.}

\vskip 0.2cm
\noindent 
In the usual real one-component case the existence of multi-soliton
solutions is a signal of integrability, but from the above we can see
that this is not necessarily true in the complex/multicomponent case.
In general it is essential that the individual solitons, from which
the multi-soliton solution is built, are each allowed to have their own
freedom of initial position and overall phase. That is, even if a
one-soliton solution is of the type \eqref{E:realr}, in the
multi-soliton case each component soliton must be allowed to have its
own parameters, including the complex coefficient(s) $c_i$.
Furthermore, under during scattering some of these parameters can
change \cite{RLH}.

Thus in practice the reduction \eqref{E:realr} trivializes the
equation and the resulting solutions are hardly of interest.
Nevertheless it seems that several recent studies have fallen into
this trap and produced no solutions with genuine multicomponent or
complex structure. This is quite evident from the proposed final
results: for example the multicomponent structure is trivialized into
a constant factor in \cite{RLD95} (see (3.15-16) or (3.20) or
(3.25-26,32-33)), \cite{RL96} (see (17) or (24) or (27)) and
\cite{N02} (see (10) or (13)), whereas the solutions are obviously
real (after the gauge has been fixed) in \cite{LW} (see Sec. III),
\cite{GKN99} (see (2,3,15,16)) and \cite{VK00} (in Sec IV $k_i,\eta_i$
are real and $H/G$ a constant). Below we will show that the reason for
this often lies in the incorrect bilinearization that was used.

\subsection{Traveling-wave solutions}
Let us now return to the one-component equations
(\ref{E:H},\ref{E:SS}) and consider their one-soliton solutions.  For
the purpose of orientation, let us first consider Hnls (\ref{E:H}).
The usual traveling-wave ansatz
\begin{equation}\label{E:An}
q(x,t)=e^{i(ax+bt+\omega)}f(x+dt+\delta),
\end{equation}
where $f$ is a real function (soliton envelope), leads to a pair of real
equations, which are compatible, if
\begin{equation}
b= a(3d- 8a^2),
\end{equation}
and in that case the solution can be parameterized as follows
\begin{equation}\label{HS1ss}
q(x,t)=e^{ia(x+(a^2-3c^2)t+\omega)}\frac{c}{\cosh[c(x+(3a^2-c^2)t+\delta)]}.
\end{equation}
We observe that there are four free real parameters: $a$ and $c$, which relate
to the size and velocity of the soliton, and $\omega,\delta$ which give the
constant complex phase and soliton position, respectively.

If we use the same ansatz (\ref{E:An}) in (\ref{E:SS}) we find that it works
only under the additional condition $a=0$, leading to
\begin{equation}\label{E:ssreal}
q(x,t)=\frac{ce^{i\omega}}{\sqrt 2\cosh[c(x-c^2t+\delta)]}.
\end{equation}
Since one parameter was lost the solution (\ref{E:ssreal}) is not
general enough.  Indeed, Sasa and Satsuma have derived a complex
traveling-wave solution to (\ref{E:SS}), which does not fit to the
usual ansatz (\ref{E:An}), but has the form \cite{SS}
\begin{equation}
q(x,t)=e^{ia(x+(a^2-3c^2)t+\omega)} \frac{2 e^\eta c (e^{2\eta}+\kappa)}
{e^{4\eta}+2e^{2\eta}+|\kappa|^2}, \quad \kappa=\frac{a}{a+ic},
\, \eta=c(x+(3a^2-c^2)t+\delta).\label{SSsol}
\end{equation}
We note that this has similar $x,t$-dependence as (\ref{HS1ss}) but the
functional form is different, also in the limit $a\to0$, i.e., $\kappa\to0$
we obtain the real limit (\ref{E:ssreal}).

It turns out that \eqref{SSsol} is still not the most general
one-soliton solution for this system, it is given by $q=G/F$, where
\begin{eqnarray}
G&=&\gamma e^\eta+\rho^*e^{\eta^*}+m
\left(\frac{\gamma}{2p^2}e^{2\eta+\eta^*}
+\frac{\rho^*}{2p^{*2}}e^{\eta+2\eta^*}\right),\label{Ohta1ss}\\
F&=&1+2\frac{|\rho|^2+|\gamma|^2}{(p+p^*)^2}e^{\eta+\eta^*}
+\frac{\rho\gamma}{2p^2}e^{2\eta}
+\frac{\rho^*\gamma^*}{2p^{*2}}e^{2\eta^*} +
\frac{|m|^2}{4|p|^4}e^{2(\eta+\eta^*)}\\
&=&1 + \tfrac12 \left| \frac{\gamma e^\eta}{p} + \frac{\rho^*
    e^{\eta^*}}{p^*} \right|^2 + \tfrac12 (|\gamma|^2+|\rho|^2)
\left|\frac{(p-p^*)e^\eta}{(p+p^*)p}\right|^2 + \left|\frac{m
    e^{2\eta}}{2 p^2}\right|^2,\nonumber\\
m&=&(|\gamma|^2p-|\rho|^2p^*)\frac{p-p^*}{(p+p^*)^2},\label{Ohtam}\\
\eta&=&px-p^3 t+\eta^{(0)},
\qquad p,\,\rho,\,\gamma \text{ and } \eta^{(0)} \text{ complex}.
\label{Ohta1sse}
\end{eqnarray}
Comparing with the original 1-soliton solution \eqref{SSsol}, we have
two extra parameters $\gamma$ and $\rho$.  By $\eta$ translation one
finds that only $\rho/\gamma$ matters, and if it vanishes we have the
usual SS-solution, so this is a genuine new parameter. This parameter
controls the oscillation, which appears not only in the carrier but
also in the envelope (but in any case $F\ge1$ so the solution is not
singular).  This solution was already obtained by Mihalache
et.~al.~\cite{MTMPT} using inverse scattering transform, below we will
derive it using the bilinear method.  It is not easy to derive such a
solution from a (complex) traveling wave ansatz, and Hirota's bilinear
method is easier to use in this case.

\subsection{Outline}
In this paper we first present in Sec.~\ref{Sec2} the bilinearizations
that work and the one-soliton solution that is obtained by the
expansion method. The detailed derivations and multi-solution
solutions are made in Sec.~\ref{Sec3}.

It is well known that soliton equations can be organized into infinite
hierarchies as described by the Sato theory \cite{JM} and that
particular equations can be obtained from these hierarchies by various
reductions. Indeed, one cannot have a full understanding of an
integrable equation before its relation to integrable hierarchies is
described. In Section \ref{Sec3} we will give the full picture by
showing that the Sasa-Satsuma equation can be obtained from the
general Sato theory as a reduction of the three-component KP
hierarchy.  The reduction can be made in two different ways producing
two different bilinearizations.  As intermediate steps of the
reduction process we get either a $(2+1)$-dimensional generalization
or a complex 2-component generalization of the Sasa-Satsuma equation.

\section{Direct bilinearization and one-soliton solutions}\label{Sec2}
One can attempt to bilinearize the generic equation
\begin{equation}
\label{E:gen}
  q_t + q_{xxx}+6|q|^2 q_x  + \beta q(|q|^2)_x=0,
\end{equation}
with the standard substitution
\begin{equation}
q=\frac{G}{F},
\end{equation}
where $F$ is taken to be real and $G$ complex.  This leads to the equation
\begin{equation}
F^2[(D_x^3+D_t)\bd GF]-\beta GF(D_x \bd G G^*)-3(D_x\bd GF)
[D_x^2\bd FF-\tfrac23(\beta+3)|G|^2]=0,
\end{equation}
which is quartic in $F,G$.  We can see that if $\beta=0$ (which is
the Hnls case) the equation splits naturally into two bilinear ones,
$(D_x^3+D_t)\bd GF=0$, and $ D_x^2\bd FF=2|G|^2$. In the general case
(that includes the SSnls equation at $\beta=3$) we could take
\begin{equation}\label{SSeq1}
D_x^2\bd FF=\tfrac23(\beta+3)|G|^2,
\end{equation}
as one of the equations, which leaves a trilinear equation
\begin{equation}\label{E:tri}
F[(D_x^3+D_t)\bd GF]-\beta G[D_x \bd G G^*]=0.
\end{equation}

One might be tempted to require that in \eqref{E:tri} the terms in
square brackets vanish separately, but this is not correct (as was
already noted in \cite{GSB97}), because it would result in more
independent equations than there are unknowns and in effect force
reduction to the real mKdV. [Clearly $D_x\bd G G^*=0\Leftrightarrow
\partial_x(G/G^*)=0$ and therefore the phase of $G$ has no
$x$-dependence, and when $G=R(x,t)e^{i\theta(t)}$ is substituted into
the remaining equation one finds that $\theta(t)$ must be constant,
i.e., \eqref{E:realr}.] As a matter of fact, this sort of {\em brute
  force bilinearization} turns out to be precisely the reason for the
trivialization of the complex/multicomponent freedom mentioned before.
Unfortunately this incorrect approach has been used quite frequently,
see, e.g., \cite{RLD95} (3.6), \cite{RL96} (9,29,44,47), \cite{PN96}
(12), \cite{P98} (26), \cite{PSM99} (21-24), \cite{GKN99} (7c),
\cite{VK00} (20), \cite{MP01} (43), \cite{MP01b} (18), \cite{N02}
(9,16), \cite{N01} (38).

The trilinear equation (\ref{E:tri}) can only be split into two
bilinear ones by introducing a new dependent variable. There are two
acceptable ways to do this, resulting in
\begin{equation}
\left\{\begin{array}{rcl}
D_x^2\bd FF&=&\tfrac23(\beta+3)|G|^2,\\
((6-\beta)D_x^3+2(\beta+3)D_t)\bd GF&=&3\beta D_x\bd HF,\\
((6-\beta)D_x^3+2(\beta+3)D_t)\bd {G^*}F&=&3\beta D_x\bd {H^*}F,\\
D_x^2\bd GF&=&-HF,\\
D_x^2\bd {G^*F}&=&-H^*F.
\end{array}\right.\label{first_bil}
\end{equation}
or
\begin{equation}
\left\{\begin{array}{rcl}
D_x^2\bd FF&=&\tfrac23(\beta+3)|G|^2,\\
(D_x^3+D_t)\bd GF&=&\beta SG,\\ 
(D_x^3+D_t)\bd {G^*}F&=&-\beta SG^*,\\ 
D_x \bd G G^*&=&SF,
\end{array}\right.
\label{alternative}\end{equation}
where the new dependent variable has been called $H$ and $S$,
respectively.  Note that $S$ is pure imaginary, $H$ complex, and that
$HG^*-H^*G=D_x\bd FS$.  These splittings are acceptable, because they
introduce equal number of new functions and new equations, and
furthermore for integrable equations and soliton solutions the new
functions turn out to be expressible in terms of polynomials of
exponentials.  Thus for any $\beta$ we can give for (\ref{E:gen}) a
bilinear form in terms of three bilinear equations for three
functions, but it should be emphasized that the fact that an equation
can be written in a bilinear form does not by itself imply that the
equation is integrable, or that the new functions $S,H$ are
$\tau$-functions, although it is the case when $\beta=3$.

The one-soliton solution can be obtained as usual by substituting the
expansions
\begin{equation}
F=1+\epsilon^2 F_2+\epsilon^4 F_4+\dots, \quad G=\epsilon
G_1+\epsilon^3 G_3+\dots
\end{equation}
accompanied with suitable ansatze $H$ or $S$, into (\ref{first_bil})
or (\ref{alternative}), and truncating at some power of the formal
expansion parameter $\epsilon$. For Hnls ($\beta=0$) the expansion can
be truncated by keeping terms up to $\epsilon^2$, but for SSnls
($\beta=3$) we must go up to $\epsilon^4$ obtaining $F,G$ as given in
(\ref{Ohta1ss}-\ref{Ohta1sse}), with the auxiliary functions
\begin{eqnarray}
S&=&(p-p^*)(|\gamma|^2-|\rho|^2)e^{\eta+\eta^*},\\
H&=&-\gamma\, p^2 e^\eta-\rho^* {p^*}^2 e^{\eta^*}
-\frac{m}2\left[\frac{\gamma {p^*}^2}{p^2}e^{2\eta+\eta^*}+
\frac{\rho^* {p}^2}{{p^*}^2}e^{\eta+2\eta^*}\right],
\end{eqnarray}
which also are polynomials of exponentials.  It is not known whether
the expansion can be truncated for any other value of $\beta$.

\section{The Sasa-Satsuma equation as a reduction of the 3-component KP
  hierarchy\label{Sec3}} We will next explain how the Sasa-Satsuma
equation and its multi-soliton solutions can be obtained from the
3-component KP hierarchy by suitable reductions. It turns out that
there are {\em two} different reduction routes leading to the
Sasa-Satsuma equation; both are two-step reductions but the
intermediate equations are different. We will first describe the
starting point (3-component KP hierarchy) and then the two kinds of
reductions.

\subsection{3-component KP hierarchy and its $\tau$-functions} 
In general, the 3-component KP hierarchy has $\tau$-functions
depending on three infinite sets of variables $\mathbf x=
x,x_2,x_3,\dots$, $\mathbf y=y,y_2,y_3,\dots$ and $\mathbf z=
z,z_2,z_3,\dots$, and is defined in terms of vector ``eigenfunctions''
$\bphi(\mathbf x)$, $\bpsi(\mathbf y)$ and $\bchi(\mathbf z)$ and
``adjoint eigenfunctions'' $\bphib (\mathbf x)$, $\bpsib (\mathbf y)$
and $\bchib (\mathbf z)$. We should emphasize that, at this point,
these six eigenfunctions are independent of one another. In general,
they are only assumed to satisfy the linear equations (for
$n=2,3,\dots$)
\begin{equation}\label{phi eqn}
    \partial_{x_n}\bphi=\partial_x^n\bphi,\quad
    -\partial_{x_n}\bphib =(-\partial_x)^n\bphib,
\end{equation}
\begin{equation}\label{psi eqn}
    \partial_{y_n}\bpsi=\partial_y^n\bpsi,\quad
    -\partial_{y_n}\bpsib =(-\partial_y)^n\bpsib,
\end{equation}
\begin{equation}\label{chi eqn}
    \partial_{z_n}\bchi=\partial_z^n\bchi,\quad
    -\partial_{z_n}\bchib =(-\partial_z)^n\bchib.
\end{equation}
Here we consider the special case in which only dependence on
$x,x_2,x_3$, $y$ and $z$ is active and so the vectors
$\bphi(x,x_2,x_3)$ and $\bphib(x,x_2,x_3)$ satisfy
\begin{equation}\label{phi eqn here}
    \partial_{x_2}\bphi=\partial_x^2\bphi,\quad\partial_{x_3}
\bphi=\partial_x^3\bphi,\quad
    \partial_{x_2}\bphib =-\partial_x^2\bphib,\quad\partial_{x_3}
\bphib =\partial_x^3\bphib,
\end{equation}
and $\bpsi(y)$, $\bchi(z)$, $\bpsib(y)$ and $\bchib(z)$ are arbitrary
vector functions of a single variable.

A potential matrix $m$ is defined by
\begin{equation}
    \partial_x m=\bphi\bphib^t,\quad
    \partial_y m=\bpsi\bpsib^t,\quad
    \partial_z m=\bchi\bchib^t,
\end{equation}
which can be  integrated to
\begin{equation}\label{mint}
    m=c+\int \bphi\bphib^t dx+
   \int \bpsi\bpsib^t dy+
   \int \bchi\bchib^t dz,
\end{equation}
where $c$ is a constant matrix.
As a consequence of \eqref{phi eqn}, we also have
\begin{equation}\label{phi x2,3}
    \partial_{x_2} m=\bphi_x\bphib^t -\bphi\bphib_x^t,\quad
    \partial_{x_3}
    m=\bphi_{xx}\bphib^t
    -\bphi_x\bphib_x^t+\bphi\bphib_{xx}^t,
\end{equation}

Now define the $\tau$-functions
\begin{equation}\label{f}
  f=\begin{vmatrix}
  m
  \end{vmatrix},
\end{equation}
\begin{equation}\label{g}
  g=\begin{vmatrix}
  m&\bphi\\
  -\bpsib^t&0
  \end{vmatrix},\quad
  \bar g =\begin{vmatrix}
  m&\bpsi\\
  -\bphib^t&0
  \end{vmatrix},
\end{equation}
and
\begin{equation}\label{h}
  h=\begin{vmatrix}
  m&\bphi\\
  -\bchib^t&0
  \end{vmatrix},\quad
  \bar h=\begin{vmatrix}
  m&\bchi\\
  -\bphib^t&0
  \end{vmatrix}.
\end{equation}

By considering Jacobi determinantal identities involving $f$, $g$,
$\bar g$, $h$ and $\bar h$ and their derivatives with respect to $x$,
$x_2$, $x_3$, $y$ and $z$, one may compile a complete list of bilinear
equations that are satisfied by these functions. The bilinear
equations given below are the only ones from this list that will
actually be used in the rest of this paper:
\begin{align}
 \label{H1} (D_x^2-D_{x_2})g\cdot f=0,\ &
\quad(D_x^2+D_{x_2})\bar g\cdot f=0,\\
 \label{H2} (D_x^2-D_{x_2})h\cdot f=0,\ & 
\quad(D_x^2+D_{x_2})\bar h\cdot f=0,\\
 \label{H3} (D_x^3+3D_xD_{x_2}-4D_{x_3})g\cdot f=0,\ &
\quad(D_x^3-3D_xD_{x_2}-4D_{x_3})\bar g\cdot f=0,\\
 \label{H4} (D_x^3+3D_xD_{x_2}-4D_{x_3})h\cdot f=0,\ &
\quad(D_x^3-3D_xD_{x_2}-4D_{x_3})\bar h\cdot f=0,\\
 \label{H5} D_yD_xf\cdot f=-2g\bar g,\ & \quad D_zD_xf\cdot f=-2h\bar h,\\
 \label{H6} D_y(D_x^2-D_{x_2})g\cdot f=0,\ & \quad D_y(D_x^2+D_{x_2})
\bar g\cdot f=0,\\
 \label{H7} D_z(D_x^2-D_{x_2})h\cdot f=0,\ & 
\quad D_z(D_x^2+D_{x_2})\bar h\cdot
 f=0.
\end{align}
As is typical for the multicomponent KP hierarchy, this set of
equations appears to be over-determined as it stands, having many more
equations than dependent variables.  But we already know that it has a
rather general set of solutions given above (even containing several
arbitrary functions of one variable).  It turns out that there exist
exactly the right number of differential relations amongst these
equations to guarantee their compatibility.  There is some freedom in
choosing the primary or independent equations, one choice is
(\ref{H1},\ref{H2},\ref{H3}a,\ref{H5}) [7 equations for 5 functions
and two dummy independent variables].  The remaining equations are
consequences from these or possibly just restrict some constants of
integration.  As an example consider (\ref{H3}b). From (\ref{H1}a) we
can determine $g_{x_2}$, from (\ref{H3}a) $g_{x_3}$ and from their
cross derivatives we get an equation for $f$. But now doing the same
computation for $\bar g$ from (\ref{H1}b) and (\ref{H3}b) we get the
{\em same} equation for $f$ and thus (\ref{H3}b) does not add
essential information.

In order to write (\ref{H1}-\ref{H7}) in nonlinear form let us first
introduce the dependent variables
\begin{equation}
q=\frac gf,\quad\bar q=\frac {\bar g}f,\quad r=\frac hf,\quad\bar
r=\frac {\bar h}f\quad\text{and}\quad\Phi=\tfrac12(\log
f)_x,\quad\Psi=\tfrac12(\log f)_{x_2}.
\end{equation}
Converting the bilinear equations into nonlinear form and eliminating
dependence on the auxiliary variable $x_2$ one obtains
\begin{equation}\left\{
\begin{array}{lc}
q_{xxx}+6q_x\Phi_{x}+3q(\Phi_{xx}+\Psi_x)-q_{x_3}&=0,\\
\bar q_{xxx}+6\bar q_x\Phi_{x}+3
\bar q(\Phi_{xx}-\Psi_x)-\bar q_{x_3}&=0,\\
r_{xxx}+6r_x\Phi_{x}+3r(\Phi_{xx}+\Psi_x)-r_{x_3}&=0,\\
\bar r_{xxx}+6\bar r_x\Phi_{x}+3
\bar r(\Phi_{xx}-\Psi_x)-\bar r_{x_3}&=0,
\end{array}\right.
\end{equation}
together with
\begin{equation}\left\{
\begin{array}{rcl}
\Phi_y=-\tfrac12 q{\bar q},&&\quad
\Phi_z=-\tfrac12 r\bar r,\\
\Psi_y=-\tfrac12(q_x{\bar q}-q{\bar q_x}),&&\quad
\Psi_z=-\tfrac12(r_x{\bar r}-r{\bar r_x}).
\end{array}\right.
\end{equation}
Although this looks superficially like a (3+1)-dimensional system, it in
fact describes a family of (2+1)-dimensional systems. The $y$- and
$z$-dependence arises in such a way that it could be replaced by single
variable corresponding to any linear combination of $y$ and $z$.

In the next sections we will describe a two stage reduction of
this system in which a calculation similar to that used above will
give the Sasa-Satsuma equation.

\subsection{First reduction, step 1}
The previous set of equations contains two dummy variables, $x_2$ and
one of $y,z$. In this reduction we will eliminate the dummy variables
by keeping just the leading terms in $x_2$ and $y-z$.

We start by considering eigenfunctions and adjoint eigenfunctions
possessing the symmetry
\begin{equation}
    \bphib(x,-x_2,x_3)=\bphi(x,x_2,x_3),\label{psichirel1}
\end{equation}
and the other eigenfunctions having pairwise identical forms:
\begin{equation}
    \bpsib(a)=\bchi(a),\quad\bchib(a)=\bpsi(a).\label{psichirel2}
\end{equation}
This reduction may be shown to be a natural generalization of the
3-component version of the $C$-reduction described in \cite{JM}.

Now we explore the consequences of this symmetry on the
$\tau$-functions. Using the independent variables $y=\xi+\eta$ and
$z=\xi-\eta$, the symmetry (\ref{psichirel2}) gives
\[
    \bpsib(y)=\bchi(\xi+\eta)=\bchi(z)\bigr|_{\eta\to-\eta},\quad
    \bchib(z)=\bpsi(\xi-\eta)=\bpsi(y)\bigr|_{\eta\to-\eta}.
\]
For the potential $m$ it is then easy to see that
\begin{equation}
    m(x,-x_2,x_3,\xi,-\eta)=m^t(x,x_2,x_3,\xi,\eta),
\end{equation}
as long as the constant matrix $c$ in (\ref{mint}) is taken to be symmetric.
Hence
\begin{equation}\label{f symm}
    f(x,-x_2,x_3,\xi,-\eta)=f(x,x_2,x_3,\xi,\eta).
\end{equation}
In a similar way,
\begin{equation}\label{g* symm}
    \bar g(x,-x_2,x_3,\xi,-\eta)=h(x,x_2,x_3,\xi,\eta),\qquad
    \bar h(x,-x_2,x_3,\xi,-\eta)=g(x,x_2,x_3,\xi,\eta).
\end{equation}

Next we consider the Taylor expansions of the eigenfunctions with
respect to $x_2$ and $\eta$ and obtain
\begin{equation}
    \bphi(x,x_2,x_3)=\bphi(x,0,x_3)+x_2\bphi_{xx}(x,0,x_3)+O(x_2^2),
\end{equation}
while the symmetry (\ref{psichirel1}) gives
\begin{equation}
    \bphib(x,x_2,x_3)=\bphi(x,0,x_3)-x_2\bphi_{xx}(x,0,x_3)+O(x_2^2).
\end{equation}
By a similar argument
\begin{equation}
\bpsi(y)=\bpsi(\xi)+O(\eta),\quad
\bchi(z)=\bchi(\xi)+O(\eta),\quad \bpsib(y)=\bchi(\xi)+O(\eta),\quad
\bchib(z)=\bpsi(\xi)+O(\eta).
\end{equation}
For the potential $m$ the expansion is
\begin{equation}
    m(x,x_2,x_3,\xi,\eta)=m(x,0,x_3,\xi,0)+
    x_2\left(\bphi_x\bphi^t-\bphi\bphi_x^t\right)(x,0,x_3)
    +O(x_2^2,\eta).
\end{equation}
For the $\tau$-functions we then get
\begin{equation}
    f=f_0(x,x_3,\xi)+O(x^2_2,\eta x_2,\eta^2),
\end{equation}
and
\begin{equation}\label{fmo}
    f_0=|m_0|,
\end{equation}
where $m_0=m(x,0,x_3,\xi,0)$ satisfies
\begin{equation}
    m_{0,x}=\bphi_0\bphi_0^t,\quad
    m_{0,x_3}=\bphi_{0,xx}\bphi_0^t-\bphi_{0,x}\bphi_{0,x}^t
+\bphi_0\bphi_{0,x}^t,\quad
    m_{0,\xi}=\bpsi\bchi^t+\bchi\bpsi^t,
\end{equation}
and $\bphi_0=\bphi(x,0,x_3)$ satisfies the single linear pde
\begin{equation}
    \partial_{x_3}\bphi_{0}=\partial_x^3\bphi_0.
\end{equation}
Also,
\begin{equation}
    g=g_0+x_2g_2+O(x_2^2,\eta),\qquad    h=h_0+x_2h_2+O(x_2^2,\eta),
\end{equation}
where
\begin{equation}\label{gh0}
    g_0=\begin{vmatrix}
        m_0&\bphi_0\\
        -\bchi^t&0
    \end{vmatrix},\quad
    h_0=\begin{vmatrix}
        m_0&\bphi_0\\
        -\bpsi^t&0
    \end{vmatrix},
\end{equation}
and
\begin{equation}\label{gh2}
    g_2=\begin{vmatrix}
        m_0&\bphi_{0,xx}\\
        -\bchi^t&0
    \end{vmatrix}-
    \begin{vmatrix}
        m_0&\bphi_0&\bphi_{0,x}\\
        -\bphi_0^t&0&0\\
        -\bchi^t&0&0
    \end{vmatrix},\quad
    h_2=\begin{vmatrix}
        m_0&\bphi_{0,xx}\\
        -\bpsi^t&0
    \end{vmatrix}-
    \begin{vmatrix}
        m_0&\bphi_0&\bphi_{0,x}\\
        -\bphi_0^t&0&0\\
        -\bpsi^t&0&0
    \end{vmatrix}.
\end{equation}
Finally, because of \eqref{g* symm}
\begin{equation}
    \bar g=h_0-x_2h_2+O(x_2^2,\eta),\qquad    \bar h=g_0-x_2g_2+O(x_2^2,\eta).
\end{equation}

The above discussion shows that, up to leading orders in $x_2$ and
$\eta$, the original five $\tau$-functions $f,\,g,\, h,\,\bar
g,\,\bar h$ can be written in terms of the five $\tau$-functions
$f_0,\,g_0,\,g_2,\,h_0,\,h_2$ depending only on $x$, $x_3$ and
$\xi$.

The final part of calculation is to identify an appropriate set of
five bilinear equations involving these five $\tau$-functions.
Applying the reduction to \eqref{H1}--\eqref{H5} gives
\begin{align}
 \label{H1C}    D_x^2g_0\cdot f_0=g_2f_0,\ &\quad D_x^2h_0\cdot f_0=h_2f_0,\\
 \label{H2C}    (D_x^3-4D_{x_3})g_0\cdot f_0=-3D_xg_2\cdot f_0,\ &
\quad(D_x^3-4D_{x_3})h_0\cdot f_0=-3D_xh_2\cdot f_0,\\
 \label{H3C}    D_\xi D_xf_0\cdot f_0&=-4g_0h_0.
\end{align}
Notice that in this reduction, \eqref{H1} and \eqref{H2} both give
\eqref{H1C}, \eqref{H3} and \eqref{H4} both give \eqref{H2C} and
the sum of the equations in \eqref{H5} gives \eqref{H3C}.

This set of bilinear equations (\ref{H1C})--(\ref{H3C}) is the
Hirota form of a $(2+1)$-dimensional Sasa-Satsuma equation. If we
define
\begin{equation}
    q=\frac{g_0}{f_0},\quad r=\frac{h_0}{f_0},\quad q_2=\frac{g_2}{f_0},\quad
    r_2=\frac{h_2}{f_0}\quad\text{ and }\quad \Phi=\tfrac12(\log f_0)_x
\end{equation}
then \eqref{H1C} give
\begin{equation}
    q_2=q_{xx}+4q\Phi_{x},\quad r_2=r_{xx}+4r\Phi_{x},
\end{equation}
\eqref{H2C} give
\begin{equation}
    q_{xxx}+12q_x\Phi_{x}-4q_{x_3}=-3q_{2,x},\quad
    r_{xxx}+12r_x\Phi_{x}-4r_{x_3}=-3r_{2,x}.
\end{equation}
and \eqref{H3C} gives
\begin{equation}
    \Phi_{\xi}=-qr.
\end{equation}
After eliminating $q_2$ and $r_2$ one obtains
\begin{equation}\label{2+1 SS}\left\{\begin{array}{rcl}
    q_{xxx}+6q_x\Phi_x+3q\Phi_{xx}&=&q_{x_3},\\
    r_{xxx}+6r_x\Phi_x+3r\Phi_{xx}&=&r_{x_3},\\
    \Phi_\xi&=&-qr.\end{array}\right.
\end{equation}
If we now set $r=-q^*$, $x_3=-t$ and use $U=\Phi_x$ we get a $(2+1)$-dimensional
Sasa-Satsuma equation
\begin{equation}\label{2+1 SS2}\left\{\begin{array}{rcl}
    q_t+q_{xxx}+6q_x U+3qU_x&=&0,\\
    U_\xi&=&(|q|^2)_x.\end{array}\right.
\end{equation}

\subsection{First reduction, step 2\label{3.3}}
In order to make a dimensional reduction from $(2+1)$- to
$(1+1)$-dimensional, we make a further rotation of coordinates
\begin{equation}\label{xXI}
    x=\tfrac12(X+\Xi),\quad \xi= \tfrac12 (X-\Xi),
\end{equation}
and then choose eigenfunctions so that the $\tau$-functions will be
independent of $\Xi$. Then both $X$ and $\Xi$ derivatives in \eqref{2+1
  SS} become $x$ derivatives and we obtain the Sasa-Satsuma equation
with two complex fields
\begin{equation}\label{complex SS}
    q_{xxx}-6q_xqr-3q(qr)_x-q_{x_3}=0,\quad
    r_{xxx}-6r_xqr-3r(qr)_x-r_{x_3}=0.
\end{equation}

In order to keep the solution structure in this dimensional reduction,
it is necessary to choose eigenfunctions $\bphi$, $\bpsi$ and $\bchi$
so that they are separable with a common dependence on $\Xi$. A
natural way to achieve this is to take
\begin{equation}\label{phi0}
    (\bphi_0)_i=\lambda_ie^{p_ix+p_i^3x_3}\to
\lambda_ie^{\tfrac12p_i\Xi}e^{\tfrac12p_iX+p_i^3x_3},
\end{equation}
\begin{equation}\label{psi}
    \psi_i=\hat\mu_ie^{-p_i\xi}\to\hat\mu_ie^{\tfrac12p_i\Xi}e^{-\tfrac12p_iX},
\end{equation}
\begin{equation}\label{chi}
    \chi_i=\hat\nu_ie^{-p_i\xi}\to\hat\nu_ie^{\tfrac12p_i\Xi}e^{-\tfrac12p_iX},
\end{equation}
where $\lambda_i,\hat\mu_i,\hat\nu_i$ and $p_i$ are constants. As a
result of this choice of eigenfunctions we have 
\begin{equation}\label{m012}
    (m_0)_{ij}=c_{ij}+e^{\tfrac12p_i\Xi}\left[{\displaystyle{\tfrac
    {\displaystyle{\lambda_i\lambda_je^{\tfrac12(p_i+p_j)X+(p^3_i+p^3_j)x_3}-
(\hat\mu_i\hat\nu_j+\hat\nu_i\hat\mu_j)e^{-\tfrac12(p_i+p_j)X}}}{{\displaystyle
    p_i+p_j}}}}\right]e^{\tfrac12p_j\Xi},\quad \text{ for }p_i+p_j\neq 0,
\end{equation}
where $c_{ij}$ are constants of integration [if $p_i+p_j=0$ we must
choose the coefficients properly so that from \eqref{mint} we get a
constant, which can then be absorbed into the $c$-matrix]. As a
consequence of the $C$-reduction, the matrix $c_{ij}$ has to be
symmetric. In order that $f_0=|m_0|$ be independent of $\Xi$ we must
have
\begin{equation}
    \prod_{i=1}^Le^{\tfrac12p_i\Xi}=\text{constant}
\end{equation}
and, for each $i,j\in\{1,\dots,L\}$,
\begin{equation}
    c_{ij}e^{-\tfrac12(p_i+p_j)\Xi}=\text{constant}.
\end{equation}
These are satisfied if and only if $ \sum_{i=1}^Lp_i=0, $ and for each
$i,j\in\{1,\dots,L\}$ either $ c_{ij}=0 \quad\text{or}\quad
p_i+p_j=0$. Consequently, we take $L=2N$ and then
\begin{equation}
p_{N+i}=-p_i,\,  \forall i=1,\dots,N, \qquad 
c_{ij}=\delta_{i+N,j} \; c_i-\delta_{i,j+N}c_j,\,
\forall i,j\in\{1,\dots,2N\}.\label{ceq}
\end{equation}

Finally, we show how to obtain solutions of the usual Sasa-Satsuma
equation \eqref{E:SS}, in which $r=-q^*$, where $^*$ stands for complex
conjugation. In order that this comes about, we must have $f_0$
real and $h_0^*=-g_0$, $h_2^*=-g_2$, and so we must impose the
relations
\begin{equation}\label{ccperm}
    \bphi^*_0=P\bphi_0,\quad \bpsi^*=P\bchi,\quad \bchi^*=P\bpsi,
\end{equation}
where $P$ is a permutation matrix. The simplest realization of these
conditions is to take $N=2M$, choose the permutation to be
\[
P=\begin{pmatrix}
0 & \mathbb I & 0 & 0\\
\mathbb I & 0 & 0 & 0\\
0 & 0 & 0 & \mathbb I\\
0 & 0 & \mathbb I & 0
\end{pmatrix},
\]
with $M\times M$ blocks, and 
\begin{equation}
    \bphi_0=\begin{pmatrix}
    \lambda_1e^{\tfrac12p_1X+p_1^3x_3}\\
    \vdots\\
    \lambda_Me^{\tfrac12p_MX+p_M^3x_3}\\
    \lambda^*_1e^{\tfrac12p^*_1X+p^*_1{}^3x_3}\\
    \vdots\\
    \lambda^*_Me^{\tfrac12p^*_MX+p^*_M{}^3x_3}\\
    0\\
    \vdots\\
    0\\
    0\\
    \vdots\\
    0
    \end{pmatrix},\quad
    \bpsi=\begin{pmatrix}
    0\\
    \vdots\\
    0\\
    0\\
    \vdots\\
    0\\
    \mu_1e^{\tfrac12p_1X}\\
    \vdots\\
    \mu_Me^{\tfrac12p_MX}\\
   \nu_1^* e^{\tfrac12p^*_1 X}\\
    \vdots\\
   \nu_M^* e^{\tfrac12p^*_M X}
    \end{pmatrix}\quad\text{and}\quad
    \bchi=\begin{pmatrix}
    0\\
    \vdots\\
    0\\
    0\\
    \vdots\\
    0\\
    \nu_1 e^{\tfrac12p_1 X}\\
    \vdots\\
   \nu_M e^{\tfrac12p_M X}\\
    \mu_1^*e^{\tfrac12p^*_1X}\\
    \vdots\\
    \mu_M^*e^{\tfrac12p^*_MX}
    \end{pmatrix},\label{vec1}
\end{equation}
where we have changed the notation for coefficients in order to
conform with \eqref{ccperm}. Since the $e^{\tfrac12p_k\Xi}$-factors in
(\ref{phi0}-\ref{m012}) will eventually cancel out with the above
choices we do not include them in these formulae, but in order to
compensate this omission we must replace $\partial_x^n\phi_0$ by
$(2\partial_X)^n\phi_0$, e.g., in \eqref{gh2}.

Taking all constants of integration $c_i=1$ in \eqref{ceq} gives the
$M$-soliton solution. In particular the one-soliton solution shown in
(\ref{Ohta1ss}-\ref{Ohta1sse}) is obtained for $M=1,\, \lambda_1=1,
\mu_1 =-\rho,\, \nu_1 =-\gamma$.

If we set $x_3=-T$ and following \eqref{xXI} replace $D_x$ and $D_\xi$
with $D_X$ in the bilinear equations (\ref{H1C}-\ref{H3C}) they become
\begin{align}
 \label{H1D}    D_X^2g_0\cdot f_0&=g_2f_0,\\
 \label{H2D}    (D_X^3+4 D_{T})g_0\cdot f_0&=-3D_Xg_2\cdot f_0,\\
 \label{H3D}    D_X^2f_0\cdot f_0&=4g_0g^*_0.
\end{align}
This is the same as (\ref{first_bil}) for $\beta=3$, if we identify
$f_0=F$, $g_0=-G$ and $g_2=H$. The multisoliton solutions are obtained
from (\ref{m012},\ref{fmo},\ref{gh0},\ref{gh2}) with \eqref{vec1}.
[But please remember that due to the simplified expressions we have
$\partial_x^n\phi_0=(2\partial_X)^n\phi_0$.]

\subsection{Second reduction, step 1}
To obtain the alternative bilinear form of SSnls we carry out the
reduction process in a different manner.  This process will take
us via a ``coupled Sasa-Satsuma equation'' as opposed to the
$(2+1)$-dimensional Sasa-Satsuma equation.

Firstly we need to introduce two new $\tau$-functions,
\begin{equation}\label{s}
s=
 \begin{vmatrix}
                 m      &\bphi&\bphi_x\\
        -{\bpsib}^t &0    &0\\
        -{\bchib}^t &0    &0
    \end{vmatrix},
    \quad
 \bar s=
 \begin{vmatrix}
                 m      &\bpsi&\bchi\\
        -{\bphib}^t &0    &0\\
        -{\bphib}_x^{t} &0    &0
    \end{vmatrix}.
\end{equation}
In addition to (\ref{H1}-\ref{H7}) we now have some further bilinear
equations satisfied by these $\tau$-functions together with the
original five $\tau$-functions (\ref{f}-\ref{h}):
\begin{align}
\label{OB3}
D_z(D_x^2-D_{x_2})g\cdot f=4s\bar h,& \qquad
D_z(D_x^2+D_{x_2})\bar g\cdot f=4\bar sh,\\
\label{OB6}
D_y(D_x^2-D_{x_2})h\cdot f=-4s\bar g,& \qquad
D_y(D_x^2+D_{x_2})\bar h\cdot
f=-4\bar sg,\\
\label{OB7}
D_xh\cdot g=sf,& \qquad
D_x\bar h\cdot\bar g=\bar sf.
\end{align}
Again these equations are not all independent.  Altogether there are 7
dependent variables $f,g,\bar g,h,\bar h,s,\bar s$ and two dummy
independent variables and therefore we need 9 independent equations.
We can take, e.g., (\ref{H1}, \ref{H2}, \ref{H3}a, \ref{H5},
\ref{OB7}), and then the other equations are consequences of these.
[In practice it is best to keep the full set at one's disposal.]

If we change variables to $\xi$ and $\eta$ using
\begin{equation}
y=\xi+\eta, \qquad z=\xi-\eta,
\end{equation}
then taking sums and differences of some of the equations, for instance
(\ref{H6}a) and (\ref{OB3}a) we get some equations containing only
$\xi$-derivatives and others containg $\eta$-derivatives. In the
following we will only use the ones containing $\xi$-derivatives, they
are
\begin{align}
\label{start2}
D_\xi(D_x^2-D_{x_2})g\cdot f=4s\bar h,& \quad
D_\xi(D_x^2+D_{x_2})\bar g\cdot f=4\bar sh,\\
D_\xi(D_x^2-D_{x_2})h\cdot f=-4s\bar g,& \quad
D_\xi(D_x^2+D_{x_2})\bar h\cdot
f=-4\bar sg,\\
\label{end2}
D_\xi D_xf\cdot f&=-2(g\bar g +h\bar h).
\end{align}
This leaves us with equations (\ref{H1}-\ref{H4}, \ref{OB7},
\ref{start2}-\ref{end2}).

At this stage we have not yet carried out a reduction. If we now do a second
change of variables
\begin{equation}
x=\tfrac12(X+\Xi),\qquad  \xi=\tfrac12(X-\Xi),
\end{equation}
we can achieve a dimensional reduction in a manner similar to the
dimensional reduction in Section \ref{3.3}, i.e., by expanding in
$\eta,\, \Xi$ and keeping only the leading terms. After also
eliminating the $x_2$ dependence, and denoting $x_3=-T$, we finally
obtain the following set of equations
\begin{align}
( D_X^3+D_T)g\cdot f=3s\bar h, &\quad
( D_X^3+D_T)\bar g\cdot f=3\bar s h,\\
( D_X^3+D_T)h\cdot f=-3s\bar g,&
\quad ( D_X^3+D_T)\bar h\cdot f=-3\bar s g,\\
D_Xh\cdot g=s f, &\quad D_X\bar h\cdot\bar g= \bar s f,\\
D_X^2f \cdot f&=-2(g\bar g+h\bar h).
\end{align}
This is a coupled Sasa-Satsuma equation with complex fields. The
nonlinear form obtained with the substitutions
\[
q=\frac gf,\quad\bar q=\frac {\bar g}f,\quad r=\frac hf,\quad\bar
r=\frac {\bar h}f,
\]
is
\begin{equation}\label{YO}\left\{
\begin{array}{rl}
q_T+
q_{XXX}-6 q_X q \bar q  -3 \bar r (q r)_X =0&,\\
\bar q_T+
\bar q_{XXX}-6  \bar q_X \bar q q
-3 r (\bar q  \bar r)_X =0&,\\
r_T+ r_{XXX}-6 r_X r \bar r  -3  \bar q (r q)_X =0&,\\
\bar r_T+
\bar r_{XXX}-6  \bar r_X \bar r r -3 q (\bar r  \bar q)_X =0&,
\end{array}\right.
\end{equation}
which was proposed already in \cite{YO}.  If we take $\bar q=-q^*,\bar
r=-r^*$ we obtain \eqref{E:sako4}.

\subsection{Second reduction, step 2}
The final reduction on this system is a reduction of $C$ type, this is
obtained as in the other bilinearization by identifying
\begin{equation}
\bar g=h, \quad
\bar h=g, \quad
\bar s =-s.
\end{equation}
This gives us the alternative bilinear form of the Sasa-Satsuma equations:
\begin{align}
( D_X^3+D_T)g\cdot f=3s g,&\quad
( D_X^3+D_T)\bar g\cdot f=-3s \bar g,\\
 D_X g\cdot \bar g&=- s f,\\
D_X^2f\cdot f&=-4 g \bar g.
\end{align}
With $f=F$, $g=-G$, $\bar g=G^*$, $s=S$  these equations
yield (\ref{alternative}) for $\beta=3$.  The solutions for this
alternative form will be the same as in the first case and the
nonlinear form of these equations is (\ref{complex SS}), with $r$
replaced by $\bar q$.  This
bilinear form of the system requires a single pure imaginary auxiliary
variable $s$, whilst the other bilinearization involves a complex
auxiliary field $h$, and consequently here we need four bilinear
equations rather than five.

\section{Conclusions}
In this paper we have shown how the Sasa-Satsuma equation fits into
the general theory as a reduction of the three-component KP hierarchy.
As a result we have obtained two possible bilinearization for SSnls,
\eqref{first_bil} and \eqref{alternative}, and formulae for
constructing multisoliton solutions, (\ref{m012}, \ref{vec1}, \ref{fmo},
\ref{gh0}, \ref{gh2}, \ref{s}).  In the reduction process we have also
obtained two intermediate equations, \eqref{2+1 SS2}, and \eqref{YO},
of which the $(2+1)$-dimensional \eqref{2+1 SS2} seems to be new.

\begin{acknowledgments}
  This work was supported in part by a grant from Academy of Finland.
  J.H. and Y.O. would like to thank Dr.~Sasa, Prof.~Satsuma and
  Dr.~Tsuchida for discussions.
\end{acknowledgments}

\end{document}